\documentclass[10pt, prd, aps, amssymb, amsmath, tightenlines, showkeys, twocolumn]{revtex4}

\usepackage{graphicx}
\usepackage[T1]{fontenc}
\usepackage[latin1]{inputenc}
\usepackage[english]{babel}
\usepackage{amsmath}
\usepackage{amsfonts}
\usepackage{amssymb}
\usepackage{dsfont}
\usepackage{color}

\def\half{{\textstyle{\frac{1}{2}}}}
\def\cP{\mathcal P}
\def\cT{\mathcal T}
\def\cC{\mathcal C}
\def\cPT{\mathcal{PT}}
\def\cCPT{\mathcal{CPT}}
\def\one{\mathds 1}
\def\half{\textstyle{\frac12}}
\newcommand\scalemath[2]{\scalebox{#1}{\mbox{\ensuremath{\displaystyle #2}}}}

\begin{document}

\title{No-signaling principle and quantum brachistochrone problem in $\cPT$-symmetric fermionic two- and four-dimensional models}

\author{Alireza Beygi}\email{beygi@thphys.uni-heidelberg.de}
\author{S. P. Klevansky}\email{spk@physik.uni-heidelberg.de}

\affiliation{Institut f\"{u}r Theoretische Physik, Universit\"{a}t
Heidelberg, Philosophenweg 12, 69120 Heidelberg, Germany\\
}

\begin{abstract}
Fermionic systems differ from bosonic ones in several ways, in particular that the time-reversal operator $\cT$ is odd, $\cT^2=-\one$. For $\cPT$-symmetric bosonic systems, the no-signaling principle and the quantum brachistochrone problem have been studied to some degree, both of them controversially. In this paper, we apply the basic methods proposed for bosonic systems \cite{R1, R21} to {\it fermionic} two- and four-dimensional $\cPT$-symmetric Hamiltonians, and obtain several surprising results: We find - in contrast to the bosonic case - that the no-signaling principle is upheld for two-dimensional fermionic Hamiltonians, however, the $\cPT$ symmetry is broken. In addition, we find that the time required for the evolution from a given initial state, the spin-up, to a given final state, the spin-down, is a constant, independent of the parameters of the Hamiltonian, under the eigenvalue constraint. That is, it cannot - as in the bosonic case - be optimized. We do, however, also find a dimensional dependence: four-dimensional $\cPT$-symmetric  fermionic Hamiltonians considered here again uphold the no-signaling principle, but it is not essential that the $\cPT$ symmetry be broken. The symmetry is, however, broken if the measure of entanglement is conserved. In the four-dimensional systems, the evolution time between orthogonal states is dependent on the parameters of the Hamiltonian, with the conclusion that it again can be optimized, and approach zero under certain circumstances. However, if we require the conservation of entanglement, the transformation time between these two states becomes the same constant as found in the two-dimensional case, which coincides with the minimum time for such a transformation to take place in the Hermitian case.

\end{abstract}


\maketitle

\section{Introduction}
\label{s1}
Since the seminal work of Bender and Boettcher \cite{BeBo}, the properties of systems with $\cPT$ symmetry have been studied extensively, and have led to important new insights \cite{Ber}. For the most part, these studies encompass bosonic systems, where time-reversal symmetry, $\cT$, is represented simply by complex conjugation.  For fermionic systems, the situation is more complicated: the fact that $\cT^2=-\one$, leads to essential differences in the formulation and the possible outcomes. One notes, for example, that if a $\cPT$-symmetric Hamiltonian $H$ describing fermions has a real eigenvalue, then $H$ has a corresponding degenerate pair of eigenvectors $\psi$ and $\cPT \psi$, which is a consequence of Kramer's theorem for conventional quantum mechanics. Non-Hermitian fermionic systems have been studied within the wider framework of pseudo-Hermiticity \cite{Mos}.

In a previous paper, we have constructed two- and four-dimensional representations of the $\cPT$- and $\cCPT$-symmetric fermionic algebras \cite{R23}, and constructed a many-body second-quantized non-Hermitian $\cPT$-symmetric Hamiltonian modelled on quantum electrodynamics, which we were able to solve exactly. Based on the knowledge gained in that work, we now  study two current problems of {\it fermionic} systems, which previously have only been discussed for bosons, notably quite controversially in the literature. These are a) The no-signaling principle, and b) The quantum brachistochrone problem. Both have a common technical feature: the time-evolution operator for a non-Hermitian $\cPT$-symmetric Hamiltonian must be considered. We discuss these problems in turn.

{\it a) The no-signaling principle.}

Lee {\it et al.~}\cite{R1} initiated the discussion of the no-signaling principle by studying a two-dimensional bosonic locally $\cPT$-symmetric Hamiltonian $H_{2\times 2}$ attributed to Alice combined with a two-dimensional Hermitian Hamiltonian attributed to Bob, the latter taken trivially to be $\one$. The combined system is then given as $H_{tot} = H_{2\times2}\otimes \one$. Both parties start out with an initial maximally entangled state, given by
$| \psi\rangle = (1/\sqrt 2)(|+_x+_x\rangle + |-_x -_x\rangle)$, where $|\pm_k\rangle$ are the eigenstates of the Pauli matrices $\sigma_k$, $k=x,y,z$. They then evaluate the no-signaling condition \cite{SP, nos},
\begin{equation}
\sum_a P(a,b|A_+,B) = \sum_a P(a,b|A_-,B) = P(b|B),
\label{eq:1}
\end{equation}
where $a$ and $b$ are the measurement outcomes of our two spacelike separated parties Alice and Bob, and $A_\pm$ and $B$ are different local measurements done by Alice and Bob on their respective sides. This condition means that the probability distribution of Bob over his measurement outcomes is unaffected by Alice's choice of measurements on her side.

The first assumption made is that a local $\cPT$-symmetric (bosonic) Hamiltonian can coexist with a Hermitian Hamiltonian. The second - and perhaps more surprising assumption - is that the authors assume that the post-measurement probabilities that must be computed in evaluating (\ref{eq:1}) should be performed within the framework of a conventional Hilbert space prescription, using a standard Dirac inner product. These authors find that if one requires the condition $\sum_a P(a,b|A_+,B) = \sum_a P(a,b|A_-,B)$ to be respected, Alice's Hamiltonian is forced to be Hermitian.  They thus conclude that the no-signaling principle is violated for all $2\times 2$ (nontrivial) $\cPT$-symmetric Hamiltonians with even time-reversal, $\cT^2=+1$. Although they do not prove it explicitly, they claim that the use of a $\cCPT$ inner product does not cure this problem.

Subsequent to this, in a detailed calculation, Japaridze {\it et al.} \cite{R2} have revisited this problem, and concluded that the calculations, redone using the $\cCPT$ inner product for the evaluation of the probabilities, in fact does preserve the no-signaling principle. In the further literature, Brody \cite{BBB} discusses the physical applicability of the claims of \cite{R1}, and demonstrates the consistency of $\cPT$-symmetric quantum mechanics with special relativity, through the proposal that the metric operator on Hilbert space is not an observable. In other words, the author claims that there is no statistical test that can be performed on the outcomes of measurements with the aim of distinguishing between  Hermiticity and $\cPT$ symmetry of a given Hamiltonian, at least for closed systems in finite dimensions.

In the work presented in this paper, we return to the ansatz of \cite{R1}, and ask the question as to how the outcomes will differ for fermionic systems. To this end, we perform calculations for both $2\times2$ and $4\times 4$  $\cPT$-symmetric fermionic matrix Hamiltonians. We arrive at the surprising results that the no-signaling principle - as discussed in the formalism of \cite{R1} - is upheld, even with the unusual calculational constraints of using the conventional Dirac inner product. In addition, we also discuss this by calculating the marginal probabilities, and also show that the measure of entanglement is conserved. However, we find that $\cPT$ symmetry is broken in the two-dimensional case, while this symmetry breaking is not essential in four dimensions unless the requirement of conservation of the entanglement is imposed: in this case, the $\cPT$ symmetry of the Hamiltonian is broken.

{\it b) The $\cPT$-symmetric fermionic quantum brachistochrone problem.}

The quantum brachistochrone problem is an attempt to find the minimal time required to transform a given initial state to a given final state in a system governed by a parametrized Hamiltonian $H$, while the difference between the largest and smallest eigenvalues is held fixed \cite{666, R21}. This has been studied by Bender {\it et al.} \cite{R21} who have chosen a (bosonic) $\cPT$-symmetric matrix Hamiltonian and studied the optimal time required to evolve a spin-up state to a spin-down one. These authors find the intriguing result that the evolution time can approach zero, provided that the elements of the Hamiltonian be extremely large.

For our case, we find a surprising result, viz. that the time to transform a spin-up state to a spin-down one, under the same eigenvalue constraint, is a constant, independent of the parameters of the Hamiltonian. This constant is the same as the optimal time for such a transformation in the Hermitian case \cite{R21}. We make the crucial observation that in the two-dimensional case, the spin-up and spin-down states are in fact orthogonal to each other with respect to the $\cCPT$ inner product. In four dimensions this is not the case, and then a dependence on the parameters of the Hamiltonian arises, so that the required time can be optimized, and be made arbitrarily small. However, if we take the conservation of the entanglement into account, the transformation time becomes the same constant as in the two-dimensional case.

This paper is structured as follows. In Sec.~\ref{ss}, we discuss the no-signaling principle, and the quantum brachistochrone problem for the $2\times 2$ $\cPT$-symmetric fermionic Hamiltonians. In Sec.~\ref{s3} both are elucidated for the $4\times4$ case. We provide some further notes on $\cPT$-symmetric quantum state discrimination in Sec.~\ref{s4}, and make some concluding remarks in Sec.~\ref{s5}.

\section{Two-dimensional model}\label{ss}
\subsection{No-signaling principle}\label{s2}
A general $\cPT$-symmetric fermionic two-dimensional Hamiltonian is described by \cite{R23},
\begin{equation}
H = \left( \begin{array}{cc} \alpha & \beta\\ \gamma & \alpha\\ \end{array} \right)
\qquad (\alpha, \, \beta, \, \gamma ~{\rm real}),
\label{1}
\end{equation}
which is self-adjoint with respect to the $\cPT$ inner product for fermions, and it commutes with $\cPT$.
\\
We recall that the fermionic $\cPT$ inner product is defined as \cite{R3},
\begin{equation}
\langle\phi|\psi\rangle_\cPT = (\cPT \phi)^T Z \psi,
\end{equation}
where the parity $\cP$, being a linear operator, can be represented by a matrix $S$ as $\cP \psi = S \psi$, and time-reversal $\cT$, being an antilinear operator, can be represented by a matrix $Z$ combined with the complex conjugation operation, i.e. $\cT \psi = Z \psi^*$. 

Alice and Bob are two spacelike separated parties,  who wish to communicate with each other without using any classical protocol. Without loss of generality, we can assume that Alice's system is governed by a special case of (\ref{1}) as
\begin{equation}
H = \left( \begin{array}{cc} 1 & \sin\alpha\\ \cos\alpha & 1\\ \end{array} \right),
\label{2}
\end{equation}
and Bob's by the identity matrix. The two parties do not interact with each other.

The eigenvalues of (\ref{2}) read
\begin{equation}
\lambda_\pm = 1 \pm \sqrt{\half\sin2\alpha},
\label{ei}
\end{equation}
with the corresponding eigenvectors
\begin{eqnarray}
|\lambda_+\rangle &=&\textstyle{ \frac{1}{\sqrt{2}}} \left( \begin{array}{c}
\sqrt[4]{\tan\alpha}\\ \sqrt[4]{\cot\alpha}\end{array}
\right), \nonumber \\
 |\lambda_-\rangle &=& \textstyle{\frac{1}{\sqrt{2}}} \left( \begin{array}{c}
\sqrt[4]{\tan\alpha}\\ -\sqrt[4]{\cot\alpha}\end{array} 
\right). \nonumber
\end{eqnarray}
\\
The eigenvalues of $H$ in (\ref{ei}) are real, provided that $\sin2\alpha > 0$. This inequality defines the region of unbroken $\cPT$ symmetry.

We note that the eigenvectors of $H$ are not orthogonal to each other with respect to the conventional Dirac inner product, however, it is easy to establish that $\langle\lambda_-|\lambda_+\rangle_\cPT = 0$.

The time-evolution operator regarding Alice's Hamiltonian can be evaluated as
\begin{equation}
U = e^{-i H t} = -\frac{2i }{\sqrt{2 \sin2\alpha}} e^{-i \pi / \omega} \left( \begin{array}{cc} 0 & \sin\alpha\\ \cos\alpha & 0\\ \end{array} \right),
\label{alice}
\end{equation}
where $\omega = \lambda_+ - \lambda_-$, and we have set $t = \pi / \omega$.

If Alice performs the measurement $\one$ with respect to the information that she wants to send to Bob, the state vector of the composite system of Alice and Bob after $t = \pi / \omega$ evolves to
\begin{equation}
|\psi_f^+\rangle = (U \one \otimes \one) |\psi\rangle,
\end{equation}
where $|\psi\rangle$ is the shared maximally entangled state described in terms of the eigenvectors of $\sigma_x$ as
\begin{equation}
|\psi\rangle = \frac{1}{\sqrt{2}} (|+_x\rangle \otimes |+_x\rangle + |-_x\rangle \otimes |-_x\rangle),
\end{equation}
and which is used by the two parties to discuss their communication protocol beforehand.

The measure of entanglement \cite{R31},
\begin{equation}
E = -tr_A (\rho_A \log\rho_A) = -tr_B (\rho_B \log\rho_B),
\label{ee}
\end{equation}
implies that $E(\psi) = 1$, where
\begin{equation}
\rho_A = \rho_B = \frac{1}{2} \left( \begin{array}{cc} 1 & 0\\ 0 & 1\\ \end{array} \right).
\end{equation}

Thus, the final state reads
\begin{equation}
|\psi_f^+\rangle = -i e^{-i \pi / \omega} \left( \begin{array}{c} 0\\ \sin\alpha\\ \cos\alpha\\ 0\\ \end{array} \right).
\label{7}
\end{equation}
We note that (\ref{7}) is normalized with regard to the conventional Dirac inner product for Hermitian quantum mechanics.

Now, if Alice performs the measurement $\sigma_x$, the final state of the composite system after $t = \pi / \omega$ becomes
\begin{equation}
|\psi_f^-\rangle = (U \sigma_x \otimes \one) |\psi\rangle,
\end{equation}
which is given explicitly as
\begin{equation}
|\psi_f^-\rangle = -i e^{-i \pi / \omega} \left( \begin{array}{c} \sin\alpha\\ 0\\ 0\\ \cos\alpha\\ \end{array} \right),
\end{equation}
where it is normalized as before.

Bob's density matrix when Alice performs the measurement $\one$ is
\begin{equation}
\rho_B^+ = Tr_A (|\psi_f^+\rangle \langle\psi_f^+|),
\end{equation}
which takes the form
\begin{equation}
\rho_B^+ = \left( \begin{array}{cc} \cos^2\alpha & 0\\ 0 & \sin^2\alpha\\ \end{array} \right).
\end{equation}
For Alice's second measurement, Bob's density matrix reads
\begin{equation}
\rho_B^- = \left( \begin{array}{cc} \sin^2\alpha & 0\\ 0 & \cos^2\alpha\\ \end{array} \right).
\end{equation}

In order for the no-signaling principle to be respected, Bob's density matrix should not be dependent on Alice's choice of measurements, that is
\begin{equation}
\rho_B^+ = \rho_B^-,
\end{equation}
which can be fulfilled if $\cos\alpha = - \sin\alpha$, implying that Alice's Hamiltonian, (\ref{2}), is still non-Hermitian and $\cPT$-symmetric. However, the symmetry is broken.

Now, if Alice and Bob measure their corresponding subsystems with the conventional quantum projectors $|\pm_y\rangle \langle\pm_y|$, we find
\begin{equation}
P(a, b | A_\pm, B) = \langle\psi_f^\pm| (|a\rangle \langle a| \otimes |b\rangle \langle b|) |\psi_f^\pm\rangle,
\end{equation}
for the joint probabilities, where $A_\pm$ correspond to the measurements, $\one$ and $\sigma_x$, performed by Alice, and $a$ and $b$ are the possible outcomes as $\pm_y$, i.e. the eigenvectors of $\sigma_y$.

The two marginal probabilities are found to be
\begin{equation}
\scalemath{0.9}{\sum_{a = \pm y} P(a, +_y | A_+, B) = \sum_{a = \pm y} P(a, +_y | A_-, B) = \frac{1}{2}}.
\end{equation}
The above calculation shows that Bob's probability distribution over his local measurement outcomes is not altered by Alice's choice of measurements on her side, that is to say, the no-signaling principle is respected.

We conclude that whether the symmetry is broken or not, Alice's Hamiltonian remains $\cPT$-symmetric without violating the no-signaling principle.

As a side remark, we note that the measure of entanglement is also conserved. Our starting point was a maximally entangled state, and at the end we still have a maximally entangled one. To see this, first we obtain the reduced density matrix of Bob, $\rho_B$. For doing so, we calculate the density matrix of the composite system after time $t = \pi/\omega$, that is
\begin{equation}
\rho = \half (|\psi_f^+\rangle \langle\psi_f^+| + |\psi_f^-\rangle \langle\psi_f^-|),
\label{rhocom}
\end{equation}
which reads
\begin{equation}
\rho = \frac{1}{2} 
\scalemath{0.85}{\left( \begin{array}{cccc} \sin^2\alpha & 0 & 0 & \sin\alpha \cos\alpha\\ 0 & \sin^2\alpha & \sin\alpha \cos\alpha & 0\\ 0 & \sin\alpha \cos\alpha & \cos^2\alpha & 0\\ \sin\alpha \cos\alpha & 0 & 0 & \cos^2\alpha\\ \end{array} \right)}.
\end{equation}
By taking the partial trace over $A$, we obtain $\rho_B$ to be
\begin{equation}
\rho_B = \frac{1}{2} \left( \begin{array}{cc} 1 & 0\\ 0 & 1\\ \end{array} \right).
\end{equation}
Equation (\ref{ee}) implies that the entanglement measure is still unity, although our time-evolution operator (\ref{alice}) is not unitary in the context of conventional Hermitian quantum mechanics.

\subsection{Quantum brachistochrone problem}

Given the initial and final states, we now investigate which $\cPT$-symmetric fermionic two-dimensional matrix Hamiltonian $H$ can achieve the transformation between these two states in the least time, provided that the difference between the largest and smallest eigenvalues of $H$ is held fixed. To approach this problem, one can determine the  optimal time for the Hamiltonian acting in the subspace spanned by the given initial and final states \cite{777}.

First, we note that the difference between the largest and the smallest eigenvalues of (\ref{1}), that is, the eigenvalue constraint $E_+ - E_- = \Omega$, reads
\begin{equation}
\Omega = 2 \sqrt{\beta \gamma}.
\end{equation}
Here $\Omega^2$ is positive if the symmetry is unbroken.

The time-evolution operator with regard to (\ref{1}) is
\begin{eqnarray}
U &=& e^{-i H t} \nonumber \\
&=& e^{-i \alpha t} \left( \begin{array}{cc} \cos\half{\Omega t} & -i \sqrt{\frac{\beta}{\gamma}} \sin\half{\Omega t}\\ -i \sqrt{\frac{\gamma}{\beta}} \sin\half{\Omega t} & \cos\half{\Omega t}\\ \end{array} \right).\nonumber \\
\end{eqnarray}

The initial state, chosen arbitrarily to be spin-up, $|\psi_0\rangle = \left( \begin{array}{c}
1\\ 0\end{array} \right)$, evolves to the final state, $|\psi_f\rangle = \left( \begin{array}{c}
a\\ b\end{array} \right)$, as
\begin{equation}
\left( \begin{array}{c}
a\\ b\end{array} \right) = U \left( \begin{array}{c}
1\\ 0\end{array} \right) = e^{-i \alpha t} \left( \begin{array}{c}
\cos\half{\Omega t}\\ -i \sqrt{\frac{\gamma}{\beta}} \sin\half{\Omega t}\end{array} \right).
\label{16}
\end{equation}
We note that the time-evolution operator preserves the $\cCPT$ norm of the initial state,  $\langle\psi_0|\psi_0\rangle_\cCPT = \langle\psi_f|\psi_f\rangle_\cCPT = \sqrt{\gamma/\beta}$, where the $\cCPT$ inner product is defined as $\langle\phi|\psi\rangle_\cCPT = (\cCPT \phi)^T Z \psi$ \cite{R3}. The $\cC$ operator reflects the sign of the $\cPT$ norm, and forces the norm of the state vectors to be positive. Thus, the Hamiltonian plays a key role in determining the operator $\cC$. For the problem at hand, its matrix representation, $K$, can be found to be
\begin{equation}
K = \left( \begin{array}{cc} 0 & \sqrt{\beta/\gamma}\\ \sqrt{\gamma/\beta} & 0\\ \end{array} \right).
\label{kkk}
\end{equation}

Now, let us assume that $a = 0$ and $b = 1$, that is, we flip the spin-up state to a spin-down one. To obtain the time required for this process, we solve for the first component of (\ref{16}), finding
\begin{equation}
t = \frac{\pi}{\Omega},
\label{brach}
\end{equation}
which is not dependent on the parameters of the Hamiltonian under the eigenvalue constraint. We also note that this constant is the minimum time for such a transformation in the Hermitian case, also called the {\it passage time} \cite{R21, R4}.

In addition, one can also show that $\textstyle{\left( \begin{array}{c} 1\\ 0\end{array} \right)}$ and $\textstyle{\left( \begin{array}{c} 0\\ 1\end{array} \right)}$ are indeed orthogonal to each other with respect to the $\cCPT$ inner product, i.e. $\langle\psi_f|\psi_0\rangle_\cCPT = 0$.

\section{Four-dimensional model}
\label{s3}

\subsection{Quantum brachistochrone problem}
A four-dimensional five-parameter Hamiltonian which satisfies all the criteria of $\cPT$-symmetric fermionic quantum mechanics, i.e. self-adjointness and invariance under $\cPT$, can be written as \cite{R3, R5},
\begin{equation}
H = \left( \begin{array}{cccc} a_0 & 0 & -C_- & -B_-\\ 0 & a_0 & -B_+ & C_+\\ C_+ & B_- & -a_0 & 0\\ B_+ & -C_- & 0 & -a_0\\ \end{array} \right),
\label{4d}
\end{equation}
where $B_\pm = b_1 \pm i b_2$, and $C_\pm = b_3 \pm i b_0$. The parameters $a_0$, $b_0$, $b_1$, $b_2$, and $b_3$ are real.

Here, the parity operator is taken to be the Dirac matrix $\gamma_0$ and the time-reversal operator as the matrix $Z$ followed by complex conjugation, where $Z=\textrm{diag}[ i\sigma_y]$. Note that, with these choices $\cP$ and $\cT$ commute, and $\cP^2$ = $\one$, and, $\cT^2$ = $-\one$. These choices for the parity and time-reversal operators are similar to those of Bjorken and Drell \cite{bd} derived in the context of coupling the Dirac electron to electromagnetic fields.

The eigenvalues of (\ref{4d}) read
\begin{equation}
E_{\pm} = \pm \sqrt{a_0^2 - b_0^2 - b_1^2 - b_2^2 - b_3^2},
\end{equation}
which are twofold degenerate. The eigenvectors corresponding to the positive energy are
\begin{widetext}
\begin{equation}
|\psi_1\rangle = \frac{i}{\sqrt{2 E_+}} \left( \begin{array}{c} \frac{\sqrt{a_0 + E_+}}{\sqrt{b_0^2 + b_1^2 + b_2^2 + b_3^2}} C_-\\ \frac{\sqrt{a_0 + E_+}}{\sqrt{b_0^2 + b_1^2 + b_2^2 + b_3^2}} B_+\\ \sqrt{a_0 - E_+}\\ 0\\ \end{array} \right),
\quad\quad\quad\quad
 |\psi_2\rangle = \cPT |\psi_1\rangle,
\end{equation}
\end{widetext}
while those corresponding to the negative energy are
\begin{widetext}
\begin{equation}
|\psi_3\rangle = \frac{i}{\sqrt{2 E_+}} \left( \begin{array}{c} \frac{\sqrt{a_0 + E_-}}{\sqrt{b_0^2 + b_1^2 + b_2^2 + b_3^2}} C_-\\ \frac{\sqrt{a_0 + E_-}}{\sqrt{b_0^2 + b_1^2 + b_2^2 + b_3^2}} B_+\\ \sqrt{a_0 - E_-}\\ 0\\ \end{array} \right), 
\quad\quad\quad\quad
|\psi_4\rangle = \cPT |\psi_3\rangle.
\end{equation}
\end{widetext}
The above degeneracy is the $\cPT$ analog of the phenomenon of Kramer's theorem in conventional Hermitian quantum mechanics, where the Hamiltonian is invariant under odd time reversal.

The eigenvalue constraint, $E_+ - E_- = \Omega$, given in terms of the parameters of the Hamiltonian, reads
$$
\Omega^2 = 4 (a_0^2 - b_0^2 - b_1^2 - b_2^2 - b_3^2),
$$
which is a positive quantity when the symmetry is unbroken, that is, the eigenvalues are real,  $a_0^2 > b_0^2 + b_1^2 + b_2^2 + b_3^2$.

Then the time-evolution operator for the Hamiltonian (\ref{4d}) is evaluated to be
\begin{widetext}
\begin{eqnarray}
U &=&e^{-iHt}\nonumber \\
 &=& \left( \begin{array}{cccc} \cos\half{\Omega t} - \frac{2i a_0 }{\Omega}\sin\half{\Omega t} & 0 & \frac{2i C_- }{\Omega}\sin\half{\Omega t} & \frac{2i B_- }{\Omega}\sin\half{\Omega t}\\ 0 & \cos\half{\Omega t} - \frac{ 2i a_0 }{\Omega}\sin\half{\Omega t} & \frac{2 iB_+ }{\Omega}\sin\half{\Omega t} & -\frac{2 iC_+ }{\Omega}\sin\half{\Omega t}\\ -\frac{2i C_+}{\Omega} \sin\half{\Omega t} & -\frac{ 2i B_- }{\Omega}\sin\half{\Omega t} & \cos\half{\Omega t} + \frac{2i a_0 }{\Omega}\sin\half{\Omega t} & 0\\ -\frac{2i B_+}{\Omega} \sin\half{\Omega t} & \frac{2i C_- }{\Omega}\sin\half{\Omega t} & 0 & \cos\half{\Omega t} + \frac{ 2i a_0 }{\Omega}\sin\half{\Omega t}\\ \end{array} \right).
\nonumber \\
\label{u}
\end{eqnarray}
\end{widetext}

The initial state, which we arbitrarily choose to be $$|\psi_0\rangle = \left( \begin{array}{c} 1\\ 0\\ 0\\ 0\end{array} \right),$$ evolves to the final state $$|\psi_f\rangle = \left( \begin{array}{c} a\\ b\\ c\\ d\end{array} \right),$$ through $U$ as
\begin{equation}
\left( \begin{array}{c}
a\\ b\\ c\\ d\end{array} \right) = U \left( \begin{array}{c}
1\\ 0\\ 0\\ 0\end{array} \right) = \left( \begin{array}{c}
\cos\half{\Omega t} - \frac{2 i a_0}{\Omega} \sin\half{\Omega t}\\ 0\\ -\frac{2 i C_+}{\Omega} \sin\half{\Omega t}\\ -\frac{2 i B_+}{\Omega} \sin\half{\Omega t}\end{array} \right).
\label{fer}
\end{equation}
To investigate whether the norm of the initial state is conserved or not, we examine the $\cCPT$ inner product for fermions as is defined in Ref. \cite{R3}, $\langle\phi|\psi\rangle_\cCPT = (\cCPT \phi)^T Z \psi$. One can also obtain the matrix representation of $\cC$ for the problem at hand as being $2H/\Omega$. Then it is easy to establish that the probability is conserved, that is, $\langle\psi_0|\psi_0\rangle_\cCPT = \langle\psi_f|\psi_f\rangle_\cCPT = 2 a_0/\Omega$.

Equation (\ref{fer}) indicates that the final state cannot be a spinor of a particle, so we consider it as corresponding to that of an antiparticle and choose, say, $a = 0$, $b = 0$, $c = 0$, and $d = 1$. (This is reminiscent of the fact that the quantum states of a particle and an antiparticle can be interchanged by applying the charge conjugation, $\cC$, parity, $\cP$, and time-reversal, $\cT$, operators.)

The first component implies that
\begin{equation}
t = \frac{2}{\Omega} \arctan(\frac{\Omega}{2 a_0}).
\label{brach1}
\end{equation}
To optimize this result over all $a_0$ positive, $t$ can approach zero as $a_0$ goes to infinity. This result requires that $|B_+|$ also be extremely large, as can be seen from the fourth component of (\ref{fer}). Thus, we can perform a spinor flip from a particle to that of an antiparticle in an arbitrarily short amount of time under the given eigenvalue constraint, provided that at least two parameters of the five-parameter $\cPT$-symmetric Hamiltonian in (\ref{4d})  be extremely large.

We recall at this point that the time for evolution between two orthogonal states in conventional quantum mechanics is limited by the uncertainty principle \cite{111}. We note, however, that the initial and final states, 
$|\psi_0\rangle$ and $|\psi_f\rangle$, are {\it not} orthogonal to each other with respect to the $\cCPT$ inner product, that is $\langle\psi_f|\psi_0\rangle_\cCPT = -2 B_+/\Omega$.

It is interesting to note that a tunable passage time could also be found numerically in the context of dissipative systems using the time-evolution operator associated with a non-Hermitian, non-$\cPT$-symmetric Hamiltonian  \cite{Assis}. This study deals with bosonic systems.

\subsection{No-signaling principle}
For simplicity, we assume that Alice's system is governed by a special case of (\ref{4d}) as
\begin{equation}
H = \left( \begin{array}{cccc} a_0 & 0 & 0 & -B_-\\ 0 & a_0 & -B_+ & 0\\ 0 & B_- & -a_0 & 0\\ B_+ & 0 & 0 & -a_0\\ \end{array} \right).
\label{aliceh}
\end{equation}

After time $t = \pi/\Omega$, the time-evolution operator with regard to Alice's system reads
\begin{equation}
U = \left( \begin{array}{cccc} -\frac{2i a_0}{\Omega} & 0 & 0 & \frac{2i B_-}{\Omega}\\ 0 & -\frac{ 2i a_0}{\Omega} & \frac{ 2i B_+}{\Omega} & 0\\ 0 & -\frac{ 2i B_-}{\Omega} & \frac{ 2 ia_0}{\Omega} & 0\\ -\frac{ 2i B_+}{\Omega} & 0 & 0 & \frac{ 2 ia_0}{\Omega}\\ \end{array} \right).
\end{equation}

As in the two-dimensional case, we assume Alice and Bob share a maximally entangled state to discuss their communication protocol beforehand:
\begin{widetext}
\begin{equation}
|\psi\rangle = \frac{1}{2} (|+_x\rangle_1 \otimes |+_x\rangle_1 + |+_x\rangle_2 \otimes |+_x\rangle_2 + |-_x\rangle_1 \otimes |-_x\rangle_1 + |-_x\rangle_2 \otimes |-_x\rangle_2),
\end{equation}
\end{widetext}
where $|\pm_x\rangle_{1, 2}$ are the eigenvectors of $\Sigma_x = \left( \begin{array}{cc} \sigma_x & 0\\ 0 & \sigma_x\\ \end{array} \right)$.

If Alice performs the measurement $\one$, after $t = \pi / \Omega$ the state vector of the composite system reads
\begin{equation}
|\psi_f^+\rangle = (U \one \otimes \one) |\psi\rangle,
\end{equation}
which becomes
\begin{equation}
|\psi_f^+\rangle = \sqrt{\frac{a_0^2 - |b|^2}{a_0^2 + |b|^2}} \left( \begin{array}{c} V^{+}\\ W^{+}\\ {W^{+}}^{'}\\ {V^{+}}^{'}\end{array} \right),
\end{equation}
where $|b|^2 = b_1^2 + b_2^2$, and
\begin{equation}
V^{+} = \left( \begin{array}{c} -\frac{i a_0}{\Omega}\\ 0\\ 0\\ \frac{i B_-}{\Omega}\end{array} \right), \qquad W^{+} = \left( \begin{array}{c} 0\\ -\frac{i a_0}{\Omega}\\ \frac{i B_+}{\Omega}\\ 0\end{array} \right).
\end{equation}
Also,
\begin{equation}
{W^{+}}^{'} = \left( \begin{array}{cccc} 0 & 0 & 0 & 0\\ 0 & 0 & 1 & 0\\ 0 & 1 & 0 & 0\\ 0 & 0 & 0 & 0\end{array} \right) {W^{+}}^{*},
\end{equation}
and
\begin{equation}
{V^{+}}^{'} = \left( \begin{array}{cccc} 0 & 0 & 0 & 1\\ 0 & 0 & 0 & 0\\ 0 & 0 & 0 & 0\\ 1 & 0 & 0 & 0\end{array} \right) {V^{+}}^{*}.
\end{equation}

Now, if Alice performs the measurement $\Sigma_x$, the final state of the composite system of Alice and Bob, after time $t = \pi/\Omega$ becomes
\begin{equation}
|\psi_f^-\rangle = (U \Sigma_x \otimes \one) |\psi\rangle,
\end{equation}
which is given explicitly as
\begin{equation}
|\psi_f^-\rangle = \sqrt{\frac{a_0^2 - |b|^2}{a_0^2 + |b|^2}} \left( \begin{array}{c} W^{-}\\ V^{-}\\ {V^{-}}^{'}\\ {W^{-}}^{'}\end{array} \right),
\end{equation}
where $W^{-}$ is obtained by replacing $B_+$ in $W^{+}$ by $B_-$ , and $V^{-}$ by replacing $B_-$ by $B_+$ in $V^{+}$.

When Alice performs the first measurement, Bob's density matrix reads
\begin{equation}
\rho_B^+ = Tr_A (|\psi_f^+\rangle \langle\psi_f^+|),
\end{equation}
where
\begin{equation}
\rho_B^+ = \frac{1}{4} \left( \begin{array}{cccc} 1 & 0 & 0 & \frac{2 a_0 B_+}{a_0^2 + |b|^2}\\ 0 & 1 & \frac{2 a_0 B_-}{a_0^2 + |b|^2} & 0\\ 0 & \frac{2 a_0 B_+}{a_0^2 + |b|^2} & 1 & 0\\ \frac{2 a_0 B_-}{a_0^2 + |b|^2} & 0 & 0 & 1\end{array} \right).
\end{equation}
For Alice's second measurement, the density matrix of Bob becomes
\begin{equation}
\rho_B^- = \frac{1}{4} \left( \begin{array}{cccc} 1 & 0 & 0 & \frac{2 a_0 B_-}{a_0^2 + |b|^2}\\ 0 & 1 & \frac{2 a_0 B_+}{a_0^2 + |b|^2} & 0\\ 0 & \frac{2 a_0 B_-}{a_0^2 + |b|^2} & 1 & 0\\ \frac{2 a_0 B_+}{a_0^2 + |b|^2} & 0 & 0 & 1\end{array} \right).
\end{equation}

The no-signaling principle is respected if $\rho_B^+ = \rho_B^-$. This requires that $B_+ = B_-$, which implies that $b_2$ must vanish. Under this constraint, the Hamiltonian that governs Alice's system, (\ref{aliceh}), is still non-Hermitian and $\cPT$-symmetric, and its eigenvalues are also real, provided that $a_0^2 > b_1^2$.

To investigate the conservation of the entanglement measure, we first construct the density matrix of the composite system as before, according to (\ref{rhocom}). Then we calculate the reduced density matrix, and by the usage of (\ref{ee}) arrive at the measure of entanglement as being
\begin{equation}
E = 1 + \frac{2 a_0 b_1}{a_0^2 + b_1^2} \log\frac{2(a_0^2 + b_1^2)}{a_0 b_1}.
\end{equation}
The measure is no longer conserved, in fact it has been increased. However, this measure can be still unity, if $a_0$ approaches zero. This also implies that the eigenvalues are no longer real, and thus that the $\cPT$ symmetry is broken. Another implication of this is that now the time required to transform between the initial and final states mentioned in the brachistochrone problem,  (\ref{brach1}), approaches $\pi/\Omega$ as $a_0$ approaches zero. This value is again, as in the two-dimensional case (\ref{brach}), the optimal time for such a transformation in the Hermitian case.

\section{A note on $\cPT$-symmetric quantum state discrimination}
\label{s4}

It is well-known that if a system is in one of two non-orthogonal quantum states, $|\psi_1\rangle$ and $|\psi_2\rangle$, it is not possible to determine with absolute certainty which state the system is in, with just one measurement \cite{222}.
This has been challenged by Bender {\it et al.} \cite{333} by exploiting the features of a non-Hermitian $\cPT$-symmetric Hamiltonian. A key point is that the inner product for such a problem is determined by the Hamiltonian at hand, that is, it is determined dynamically. Thus, it is possible to introduce a Hamiltonian in such a way that relative to its inner product the two states $|\psi_1\rangle$ and $|\psi_2\rangle$ become orthogonal.

The general $\cPT$-symmetric Hamiltonian which they consider is built on the assumption that the time-reversal operator is just complex conjugation, and as a result of this, they arrive at a \textit{complex} Hamiltonian. And, they conclude that this ability to distinguish between a pair of non-orthogonal states with a single measurement is due to the complex degrees of freedom made available by $\cPT$ symmetry.

We show that their results are still valid for the fermionic case for which it turns out that the non-Hermitian $\cPT$-symmetric Hamiltonian is real, see (\ref{1}).

First, we consider the two states, $|\psi_1\rangle$ and $|\psi_2\rangle$ on the Bloch sphere, that are separated by the angular distance $2 \epsilon$ as
\begin{equation}
|\psi_1\rangle = \left( \begin{array}{c} \cos\frac{\theta}{2}\\ e^{i \phi} \sin\frac{\theta}{2}\end{array} \right), \, |\psi_2\rangle = \left( \begin{array}{c} \cos(\frac{\theta}{2} + \epsilon)\\ e^{i \phi} \sin(\frac{\theta}{2} + \epsilon)\end{array} \right).
\end{equation}
For definiteness, we choose $\phi = \pi$, and $\theta = 2 \pi/3 - \epsilon$.

These two states are not orthogonal in the conventional sense, i.e. $\langle\psi_1|\psi_2\rangle \neq 0$.
Now, by considering (\ref{1}), (\ref{kkk}), and the inner product $\langle\psi_i|\psi_f\rangle_\cCPT = (\cCPT \psi_i)^T Z \psi_f$, we can construct the bra vector corresponding to $|\psi_1\rangle$ as being
\begin{equation}
\langle\psi_1|_\cCPT = \left( \begin{array}{c} \sqrt{\frac{\gamma}{\beta}} \cos(\frac{\pi}{3} - \frac{\epsilon}{2})\\ -\sqrt{\frac{\beta}{\gamma}} \sin(\frac{\pi}{3} - \frac{\epsilon}{2})\end{array} \right)^T.
\end{equation}
Then we require that $\langle\psi_1|\psi_2\rangle_\cCPT$ vanishes, which results in the condition that
\begin{equation}
\tan^2\frac{\epsilon}{2} = \frac{\gamma + 3 \beta}{3 \gamma + \beta}.
\end{equation}

Now, to distinguish between the two states, we need the projection operators
\begin{equation}
|\psi_1\rangle \langle\psi_1|_\cCPT \quad \textrm{and}\quad |\psi_2\rangle \langle\psi_2|_\cCPT.
\end{equation}
Thus, by applying one of these projection measurements, we can distinguish between the states $|\psi_1\rangle$ and $|\psi_2\rangle$ with absolute certainty.

\section{Brief concluding remarks}
\label{s5}
In this paper, we have applied the procedures suggested by \cite{R1, R21} for studying the no-signaling principle, and the quantum brachistochrone problem in $\cPT$-symmetric fermionic two- and four-dimensional models. The results show several interesting properties. Firstly, a dimensional dependence emerges. For the quantum brachistochrone problem, the time required to transform a spin-up state to a spin-down state in the two-dimensional case - unlike its bosonic counterpart - shows no dependence on the parameters of the Hamiltonian, and it is a constant under the eigenvalue constraint, where this constant coincides with the minimum time for such a transformation in the Hermitian case. A parameter dependence, however, re-emerges as a feature of the analysis of the four-dimensional system, and it can approach zero provided that some parameters of the Hamiltonian be extremely large.  In this case, however, one again recovers the same constant for the transformation time as in the two-dimensional case by taking the conservation of entanglement into account. In general, the brachistochrone itself may be related to the orthogonality or alignment of the initial and final states within the chosen theory.  Secondly, the no-signaling principle is upheld in the two-dimensional system, with the caveat that $\cPT$ symmetry is broken. In four dimensions, however, it is again upheld, but $\cPT$ symmetry is only broken if the conservation of entanglement is enforced.

\end{document}